\documentclass[12pt,onecolumn]{IEEEtran}

\usepackage[pdftex]{graphicx}
\usepackage{amsmath}
\usepackage{amssymb }
\usepackage{algorithm}
\usepackage{authblk}

\ifCLASSOPTIONcompsoc
\else
\fi
\ifCLASSINFOpdf
\graphicspath{{Figures/}}
\else
\fi
\begin{document}


\title{Low Complexity Power Allocation Schemes in Regenerative Multi-user Relay Networks\footnote{A part of this work  was submitted to the IEEE International
Conference on Communications (ICC), Miami, FL, USA, Jun.10-Jun.14, 2014. In this paper, we provide results with different low complexity power allocation schemes in addition to the ones proposed in \cite{OwnWork}.}}
\author[1]{\rm Arvind Chakrapani}
\author[2]{\rm Robert Malaney}
\author[2]{\rm Jinhong Yuan}
\affil[1]{Qualcomm Flarion Technologies, Bridgewater, New Jersey, USA}
\affil[1 ]{ achakrap@qti.qualcomm.com}
\affil[2]{School of Electrical Engineering and Telecommunication, University of New South Wales, Sydney, Australia}
\affil[ 2]{ {\{r.malaney,jinhong.yaun \}@unsw.edu.au}}

\maketitle

\begin{abstract}
In relay assisted wireless communications, the multi-source, single relay and single destination system (an $M$-1-1 system) is of growing importance,  due to the increased demand for higher network throughput and connectivity. Previously, power allocation in $M$-1-1 systems have assumed availability of instantaneous channel state information (CSI), which is rather idealistic. In this paper we consider an $M$-1-1 Decode-and-Forward (DF), Full-Duplex, orthogonal frequencey division multiple access (OFDMA) based relay system with statistical-CSI and analyze the achievable rate $R$ of such a system. We show how $R$ can only be maximized by numerical power allocation schemes which has a high-complexity of order $\mathcal O(M^3)$. By introducing a rational approximation in the achievable rate analysis, we develop two low-complexity power allocation schemes that can obtain a system achievable rate very close to the maximum $R$. Most importantly, we show that the complexity of our power allocation schemes is of order $\mathcal O(M\log M)$. We then show how our power allocation schemes are suitable for a multi-user relay system, where either the priority is to maximize system throughput, or where lower computations in power allocation scheme are essential. The work we present in this paper will be of value to the design and implementation of  real-time multi-user relay systems operating under realistic channel conditions.

\end{abstract}
\section{Introduction}

Recently there has been significant interest from both academia and industry in the concept of cooperative relaying in infrastructure based broadband wireless access for 4G networks, e.g., 802.16j - Mobile Multihop Relay (MMR) specification \cite{80216j}. In relay assisted communication, relay stations (RS), either fixed or mobile, are introduced to increase the capacity (for both uplink and downlink) or connectivity among mobile sources (MS). The 802.16j MMR standard, specifies two modes of relaying techniques. One is the transparent relaying mode, where relays are used to increase capacity of MS who are within the range of the Base Station (BS). The other is the non-transparent relaying mode, where relays are deployed mainly to increase the coverage area of the BS. In this work, we will focus on the uplink of a cellular system with transparent mode of relaying.

In the transparent mode, the relay is used to enhance the throughput of each source. The relaying can employ either the amplify and forward (AF) or the DF strategy. It was shown in \cite{Aria} that DF provides a higher achievable rate relative to AF at low signal-to-noise ratio (SNR). Also, the relay can operate in either full-duplex (sources and relay transmit simultaneously) or half-duplex mode (sources and relay transmit during different time slots). Note that half-duplex mode can be implemented with a single antenna whereas full-duplex mode will require additional antennas for self-intereference cancellation. It has been previously shown (e.g., \cite{CompFD}) that the full-duplex mode of relaying is spectrally more efficient than the half-duplex mode. A general precipt so far has been that the practical implementation of full-duplex mode is often not possible due to large difference in transmit and receive power at the relay.

However, recently there have been significant works both from academics (e.g., \cite{FeaseFD}-\cite{OutageRice}]) and industry (e.g., \cite{IndusFD1}-\cite{IndusFD3}) regarding the feasibility of a full-duplex relay system with DF strategy. In fact, the works in \cite{OffShelfFD} and \cite{MobicomFD} showed that a practical full-duplex system can be built using off-the-shelf hardware. Antenna seperation  with Analog/Digital Cancellation techniques was used in \cite{OffShelfFD} and the experiments showed that the full-duplex mode can be practically implemented. A novel self intereference cancellation technique was used in \cite{MobicomFD} and a working prototype was developed, which achieved median performance that was within 8$\%$ of an ideal full-duplexing system. Also, in \cite{CoopBook} a transmission policy based on block Markov encoding for a DF full-duplex relay system was described. The above works have clearly demonstrated that building a practical DF full-duplex relay system without introducing significant latency into the transmission, is indeed possible.  In this paper, we consider a DF full-duplex relay system with perfect self-interference cancellation.

In the transparent mode, the single source, relay and destination form an 1-1-1 system. The 1-1-1 system has been well investigated over the years (e.g. \cite{Cover}-\cite{Ref1rec} and the references therein) and various studies have focused on several performance aspects, including achievable rates \cite{Gastpar}, outage probabilities \cite{Outage}\cite{OutageRice}\cite{OutageInter} and power allocation \cite{Outage}\cite{OutageInter}\cite{HostMadsen}. Recently the performance of an $M$-1-1 system has gained much attention \cite{Multiuser}\cite{PhanKT}\cite{Lalitha}. Resource allocation and relay selection in a multi-user OFDMA based system was studied in \cite{Multiuser}, assuming access to full-CSI. The power allocation scheme for a multi-source AF relay system to maximize the network throughput was investigated in \cite{PhanKT}. The multi-source achievable rate and power allocation for a half-duplex relay system was investigated for a multiple access relay channel in \cite{Lalitha}, assuming availability of full-CSI at the relay.

However, none of the works have investigated practical power allocation schemes in an $M$-1-1 DF relay system in Rayleigh fading environment, when \emph{no} instantaneous CSI is available, or where only statistical information of the channel state (i.e., statistical CSI) is available. This is mainly due to the complicated nature of the throughput analysis and as such only numerical methods of optimal power allocation can be employed, which have a complexity of order $\mathcal O(M^3)$ \cite{optbook}. Such complexity renders them infeasible for implementation in real-time systems, especially when the number of users in the system increase. Note that in a 4G network, the number of users $M$ in the network is typically large \cite{80216j} and efficient power allocation at the relay will be lead to significant increase in the system throughput. In this paper, we develop two low-complexity power allocation schemes (of different computational speed) at the relay for an $M$-1-1 system with statistical-CSI. Let $R_{PAS-0}$ be the maximum achievable rate of the system obtained with an optimal power allocation scheme (we denote this as PAS-0) at the relay. We show how our power allocation schemes can obtain a system achievable rate close to $R_{PAS-0}$, and show that the complexity of our power allocation schemes is of order $\mathcal O(M\log M)$.

Our contributions reported in this paper are as follows. First, we analyze the achievable rate $R$ an $M$-1-1 system with statistical-CSI. Second (and the key contribution in this paper), is that we introduce a rational approximation in the achievable rate analysis, which helps us develop low-complexity power allocation schemes that can obtain a system throughput close to $R_{PAS-0}$. Third, using our rational approximation, we develop two low-complexity power allocation schemes (of varying computational speed) at the relay. More specifically, we develop the following two power allocation schemes.

$\bullet$ We develop a Lagrangian-based power allocation scheme (we denote this as PAS-1) that obtains an achievable rate $R_{PAS-1}$, which is approximately equal to $R_{PAS-0}$ for all practical purposes. We show through analysis and simulations that PAS-1 has negligible loss in throughput compared to PAS-0. Most importantly we show that the complexity of the PAS-1 algorithm is of order $\mathcal O(M\log M)$.

$\bullet$ Utilizing the results from the PAS-1 algorithm, we develop a second power allocation scheme (we denote this as PAS-2) which delivers a system achievable rate within $\approx 5-10\%$ of $R_{PAS-0}$ and requires lower computations compared to PAS-1 and is free of logarithmic and cube root operations.

The paper is organized as follows. In section \ref{statCSI}, we analyze the achievable rate $R$ of an $M$-1-1 system with statistical-CSI. In section \ref{statPower}, we provide the approximations required for our new power allocation schemes. In section \ref{PAS-1}, we develop a Lagrangian-based power allocation scheme (PAS-1) that obtains a system achievable rate approximately equal to $R_{PAS-0}$ for all practical purposes. In section~\ref{nopas} we develop the second power allocation scheme (PAS-2), which provides a system achievable rate within $\approx 5-10\%$ of $R_{PAS-0}$ and requires lower computations compared to the PAS-1 algorithm. In section \ref{complexity}, we discuss the computational complexity of our two power allocation schemes. In section~\ref{StatanalSimRes} we provide analytical and simulations results. Finally, in section \ref{StatConcls}, we draw conclusions.

\section{System Model}

\begin{figure}[h]
\par
\begin{center}
{\includegraphics[width=0.42\textwidth]{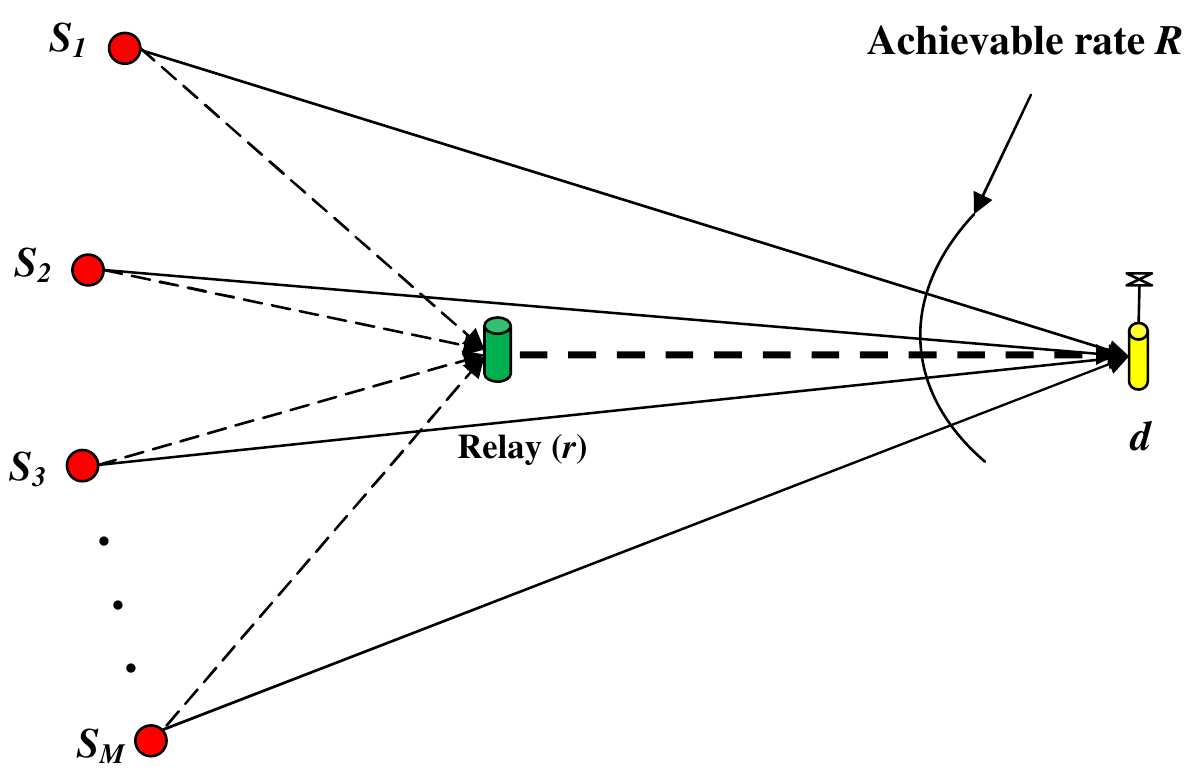}}
\end{center}
\caption{Multi-source $S_{m}, m = \{1,...,M\}$, with single relay and destination ($M$-1-1) system.}
\label{multiuserRelay}
\end{figure}

\begin{figure}[h]
\par
\begin{center}
{\includegraphics[width=0.42\textwidth]{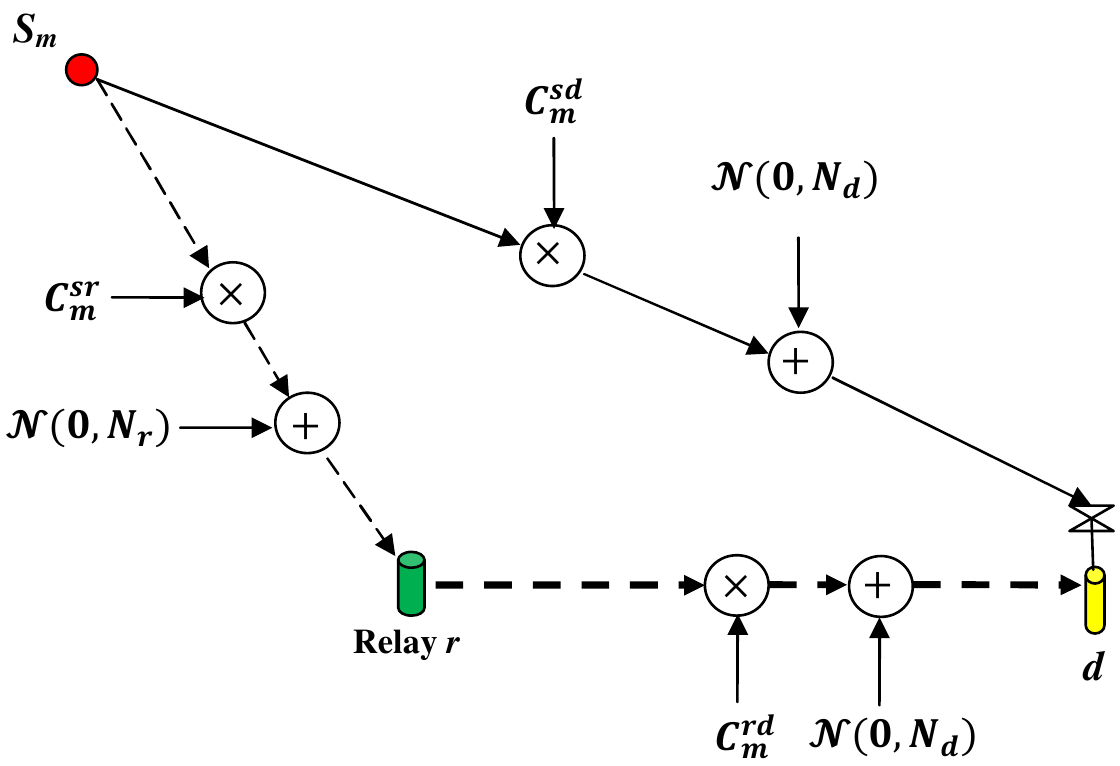}}
\end{center}
\caption{The $m$th source $S_{m}$, relay and destination form a 1-1-1 triangle model.}
\label{triangle}
\end{figure}

Consider a multi-source relay system shown in Fig.~\ref{multiuserRelay}. Sources $S_{m}, m\in \left\{1,...M \right\}$ transmit their information to the destination $d$ simultaneously with the help of a full-duplex relay $r$. A bin indexing scheme as in \cite{Gastpar} was assumed to transmit information and parity bits. The conventional DF relaying with orthogonal transmission\footnote{Here, by orthogonal transmission we mean there is no interference at the destination due to transmissions from multiple sources and relay.} through OFDMA is assumed. With OFDMA, the $m$th source $S_{m}$ transmits its messages in the frequency bands $f_{m}$, the relay $r$ receives and transmits at frequencies $f_{1},...,f_{M}$, respectively. Note that one antenna is sufficient at the relay for transmitting/receiving an OFDMA signal with $M$ sub channels.  The destination receives signals at these $M$ orthogonal frequency bands. With these constraints, the multi-source system can be viewed as $M$ independent parallel 1 - 1 - 1 triangle systems, one of which is shown in Fig.~\ref{triangle}. All channels are assumed to undergo Rayleigh fading and are corrupted with Additive White Gaussian Noise (AWGN). The source $S_{m}$ transmits in the frequency bands $f_{m}$ to the destination.
The source $S_m$ begins by encoding a $q$-bit message Qm into a codeword of length $n$ ($k < n$). The codeword is then divided into $B$ blocks of length $nr_c$ bits each, where $r_c$ is channel coding rate at the encoder ($r_c \le 1$). The coding rate $r_c$ specifies how much redundancy is transmitted with every message bit. For a $q$-bit message, $q/r$ bits are transmitted in $B$ + 1 blocks. The codeword is encoded into $c$ symbols $x_{1}[1],...,x_{1}[c]$ and transmitted over the channel, under the power constraint $\left|\frac{1}{c}\sum_{j=1}^{c}x_{1}[j]^2 \right| \le P_{s}$, where $P_{s}$ is the maximum transmit power available at each source. The relay decodes and forwards a new block $x_{2}[j]$ to aid the communication between source and destination. $x_{2}[j]$ is also encoded into $c$ symbols subject to the power constraint $\left | \frac{1}{c}\sum_{j=1}^{c}x_{2}[j]^2 \right| \le P_{m}$, where $P_{m}$ is the power allocated by the relay for transmitting the $m$th source signal. The received signal at the relay $y_{rm}$ and the destination $y_{dm}$ are given by,
\begin{eqnarray}\label{receivedSig}
y_{rm} &=&C_{m}^{sr}x_{1}[j]+n_{r} \\
y_{dm} &=&C_{m}^{sd}x_{1}[j]+C_{m}^{rd}x_{2}[j]+n_{d},
\end{eqnarray}
where $ C_{m}^{sr}$, $ C_{m}^{sr}$ and $ C_{m}^{sr}$ represent the channel gains between $S_{m}$ to $r$ (denoted as S-R), $S_{m}$ to $d$ (denoted as S-D), and $r$ to $d$ (denoted as R-D), respectively. Here, $n_{r}$ and $n_{d}$ are independent AWGN's with zero mean and unit variance. We consider a propagation model\footnote{Note that, here we have ignored the effect of shadowing on the channel gain for simplification. Including the shadowing component would scale down the achievable rate by a constant factor, but does not add any further insights.} as in \cite{CoopBook} and let,
$ C_{m}^{sr} = \frac{\left | h_{m}^{sr} \right |^2}{(d_{m}^{sr})^{\alpha}N_{r}}$, $ C_{m}^{sd} = \frac{\left | h_{m}^{sd} \right |^2}{(d_{m}^{sd})^{\alpha}N_{d}}$ and $ C_{m}^{rd} = \frac{\left | h_{m}^{rd} \right |^2}{(d_{m}^{rd})^{\alpha}N_{d}}$, where $h_{m}^{sr}$, $h_{m}^{sd}$ and $h_{m}^{rd}$ are complex fading random variables for channels between $S_{m}$ to $r$, $S_{m}$ to $d$, and $r$ to $d$, respectively. $N_{r}$ and $N_{d}$ are the noise spectral densities at the relay and at the destination respectively.  Here, $d_{m}^{sd}$, $d_{m}^{sr}$ and $d_{m}^{rd}$ represent the normalized distances between S-D, S-R and R-D respectively. Note that the distances are normalized with respect to a reference distance of $d_{0} = 1$ unit. Here, $\alpha$ represents the pathloss exponent. For a Rayleigh channel, the real and imaginary parts of the complex fading variables are Gaussian distributed having zero mean and variance 1/2.


\subsection{Problem Statement} Consider an $M$-1-1 system described above. Let the relay have a maximum total transmit power of $P_{r}$ and the transmit power $P_{s}$ at each source be fixed. We investigate the following two problems. $1)$ What is the achievable rate of the $M$-1-1 DF relay system when all channels undergo Rayleigh fading? $2)$ Let the relay allocate power among $M$ sub-channels as $\{P_{1},...,P_{M}\}$, such that $\sum_{m=1}^{M}{P_{m}} = P_{r} $. What is the power allocation vector $\{P_{1},...,P_{M}\}$ at the relay which obtains the achievable rate $R$?

\section{Achievable Rate with Statistical-CSI}\label{statCSI}

The $m$th source, the relay and the destination, form a 1-1-1 system as shown in Fig.~\ref{triangle}. The instantaneous achievable rate for such a 1-1-1 system can be expressed as (\cite{CoopBook}, Equation~(4.33)),
\begin{equation}\label{achRate2}
R_{m}^{i} = \min\left\{\log\left( 1+\frac{\left | h_{m}^{sr} \right |^2P_{s}}{(d_{m}^{sr})^{\alpha}N_{r}}\right),\log\left ( 1+ \frac{\left | h_{m}^{sd} \right |^2P_{s}}{(d_{m}^{sd})^{\alpha}N_{d}} + \frac{\left | h_{m}^{rd} \right |^2P_{m}}{(d_{m}^{rd})^{\alpha}N_{d}}\right) \right \}.
\end{equation}
Note that \eqref{achRate2} is valid (see \cite{CoopBook}, section 4.2.5) only for a fading channel where the phase is uniformly distributed over $[0,2\pi)$ (e.g., Rayleigh Fading channel), i.e., there is no correlation between the relay signal and the source signal. Note also that throughout this paper $\log(\cdot)$ represents logarithm to base 2. The achievable rate of an 1-1-1 system with averaged over all channel fading states (with Rayleigh distribution), i.e., with statistical-CSI is given by~\cite{ErgCap},
$R_{m}~=~\min\left \{ R_{1m}, R_{2m} \right \},$
\begin{eqnarray}\label{expr1}
\textrm{where,}\;\;\;\;R_{1m} = \log(e)\left [ \exp\left ( \frac{k_{m}^{sr}}{P_{s}} \right )E_{1}\left ( \frac{k_{m}^{sr}}{P_{s}} \right ) \right],
\end{eqnarray}
\begin{equation}\label{expr2}
\textrm{and}\;\;\;\;\;\;\;\;R_{2m} = \frac{\log(e)\left [P_{m}k_{m}^{sd} \exp\left ( \frac{k_{m}^{rd}}{P_{m}} \right )E_{1}\left ( \frac{k_{m}^{rd}}{P_{m}} \right ) - P_{s}k_{m}^{rd} \exp\left ( \frac{k_{m}^{sd}}{P_{s}} \right )E_{1}\left ( \frac{k_{m}^{sd}}{P_{s}} \right )\right]}{\left ( P_{m}k_{m}^{sd}-P_{s}k_{m}^{rd} \right )},
\end{equation}
where, $k_{m}^{sr} = (d_{m}^{sr})^{\alpha}N_{r}$, $k_{m}^{sd} = (d_{m}^{sd})^{\alpha}N_{d}$ and $k_{m}^{rd} = (d_{m}^{rd})^{\alpha}N_{d}$ and where, $E_{1}(\cdot)$ is the exponential integral defined as $E_{1}(x) = \int_{1}^{\infty }\frac{e^{-xt}}{t}dt, (x>0)$ and $e = \exp(1) \approx 2.7183$. It will be useful to rewrite \eqref{expr2} as, $R_{2m} = R_{2m}^{+} + R_{2m}^{-}$,
where,
\begin{equation}\label{barR2m+}
R_{2m}^{+} = \log(e)\left [ \frac{\exp\left ( \frac{k_{m}^{rd}}{P_{m}}\right )E_{1}\left ( \frac{k_{m}^{rd}}{P_{m}} \right ) - \exp\left ( \frac{k_{m}^{sd}}{P_{s}}\right )E_{1}\left ( \frac{k_{m}^{sd}}{P_{s}} \right )}{\left(1- \frac{P_{s}k_{m}^{rd}}{P_{m}k_{m}^{sd}}\right)} \right ]
\end{equation}
\begin{equation}\label{barR2m-}
\textrm{and}\;\;\;\;R_{2m}^{-} = \log(e)\left[\exp\left ( \frac{k_{m}^{sd}}{P_{s}}\right )E_{1}\left ( \frac{k_{m}^{sd}}{P_{s}} \right )\right].
\end{equation}
Since the transmissions from $M$ sources are non-interfering at the destination, an $M$-1-1 system can be considered as independent 1-1-1 systems. The achievable rate for the whole $M$-1-1 system can then be written as,
\begin{equation}
R = \sum_{m =1}^{M}\left(\min\left\{ R_{1m}, R_{2m}^{+}+ R_{2m}^{-} \right \}\right).
\end{equation}
Note that an important assumption in deriving \eqref{achRate2} (see \cite{CoopBook}), is that the S-R rate is always greater than the sum rates of S-D and R-D (i.e., the relay is able to decode the source signal). When the relay is not able to decode the source signal, the model assumption is that (e.g., \cite{Outage}\cite{OutageRice}\cite{CoopBook}) either the source is far from both the relay and destination or the source is closer to destination than the relay. In both scenarios, zero power is allocated (i.e., $P_m$ = 0 for the $m$th source) by the relay (using our proposed power allocation scheme). This acts as an admission control mechanism, where only the sources with higher SNR between themselves and the relay are admitted into the system (or allocated power at the relay). Such a scheme is efficient in avoiding wastage of power at the relay, by only admitting sources into the system whose S-R channel SNR is good (so that the relay is able to decode).

Therefore the rate $R_{1m} > R_{2m}^{+}+ R_{2m}^{-}$ $\forall m$. The achievable rate for an $M$-1-1 DF system with Rayleigh fading is then given by,
\begin{equation}\label{R_summation}
R =\min\left \{ \sum_{m =1}^{M}R_{1m}, \sum_{m =1}^{M} R_{2m}^{+}+ \sum_{m =1}^{M}R_{2m}^{-} \right \},
\end{equation}
which can be simplified as $R = \sum_{m =1}^{M} R_{2m} =  \sum_{m =1}^{M} R_{2m}^{+}+ \sum_{m =1}^{M}R_{2m}^{-}$.
Note that when an optimal power allocation scheme is found at the relay (i.e., the optimal vector $P_{1},...,P_{m}$) the achievable rate in \eqref{R_summation} is maximized. In the following section we will develop low-complexity power allocation schemes at the relay which delivers a system throughput close to the maximum $R$ (denoted as $R_{PAS-0}$).

Note that the channel gains between S-R can be measured by the relay and therefore instantaneous CSI for S-R channels can be obtained at the relay. However, the instantaneous CSI for the channels between the S-D along and the channel between the R-D is not known at the relay. Assuming availability of instantaneous CSI of all channels via feedback with zero-delay is no that practical. Therefore, in our system model, we have assumed availability of only statistical CSI for all channels, which is a more realistic setting. However, considering availability of instantaneous CSI for S-R links and availability of statistical CSI between S-D and R-D links forms a different hybrid system model. It is important to note that, our proposed power allocation scheme (in the following section) can be easily extended to the such an hybrid system model. This is apparent from (3), where we need to integrate only the sum rate $R_{2m}$ over all channel states (statistical CSI) and all other following results obtained are still applicable.

\section{Power Allocation Schemes at the Relay}\label{statPower}

In this section we investigate our power allocation schemes at the relay which obtains a system achievable rate close to $R_{PAS-0}$ (maximum $R$). Due to the minimization term in \eqref{R_summation}, the second term $\sum_{m =1}^{M} R_{2m}^{+} + \sum_{m =1}^{M}R_{2m}^{-}$ of this equation should be always less than the first term $\sum_{m =1}^{M}R_{1m}$, for the power allocation at the relay to be efficient. Therefore, any power allocation scheme at the relay must be under the constraint of $\sum_{m =1}^{M}R_{2m}^{+} \le   \sum_{m =1}^{M}R_{1m} - \sum_{m =1}^{M} R_{2m}^{-}.$
The relay has a total available power of $P_{r}$ and needs to allocate the power among $M$ users in order to maximize $R$ (i.e., to obtain $R_{PAS-0}$). Obtaining $R_{PAS-0}$ is equivalent to the maximization of $R_{2m}^{+}$. The power allocation vector $P_{1},...,P_{m}$ can be obtained by solving the following convex optimization problem with the constraints listed below.
\begin{equation}\label{maxeq}
\text{max} \{R_{2m}^{+}\} =\underset{P_{1},...,P_{m}}{\text{max}}\sum_{m =1}^{M} \left (\log(e)\left [ \frac{\exp\left ( \frac{k_{m}^{rd}}{P_{m}}\right )E_{1}\left ( \frac{k_{m}^{rd}}{P_{m}} \right ) - \exp\left ( \frac{k_{m}^{sd}}{P_{s}}\right )E_{1}\left ( \frac{k_{m}^{sd}}{P_{s}} \right )}{\left(1- \frac{P_{s}k_{m}^{rd}}{P_{m}k_{m}^{sd}}\right)} \right ]\right)
\end{equation}
subject to,
\begin{enumerate}\label{constraints}
\item $\sum_{m = 1}^{M}P_{m} = P_{r}$,
\item $P_{m} \ge 0, m = 1,\cdots ,M$,
\item $R_{2m}^{+} - R_{1m} + R_{2m}^{-} \le 0, m = 1,\cdots,M $.
\end{enumerate}
Since our objective is to maximize the throughput in the network, we need to allocate all the power available at the relay and the first constraint \eqref{maxeq} is required. Note that in our system model the relay allocates lower power to sources which are closer the destination than the relay. This is because the throughput increase with the help of a relay will not be significant for sources close to the destination. Note also that the instantaneous CSI is not required for the power allocation at the relay to maximize the achievable rate $R$. The optimization problem in \eqref{maxeq} can be solved using numerical optimization tools. However, in practical relay systems, numerical search algorithm may not be practical. Even the most efficient optimization search algorithms are known to have complexity of the order 
$\mathcal O(M^3)$ (e.g. interior point method \cite{optbook}). The complexity of such algorithms scales with the number of users, making them intractable. We therefore develop two low-complexity power allocation schemes (PAS-1 and PAS-2) which can be easily implemented in a real-time system.

Due to the non-linear product $\exp\left(\frac{k_m^{rd}}{P_m}\right)E_1\left(\frac{k_m^{rd}}{P_{m}}\right)$ in \eqref{maxeq}, the classical water-filling (CWF) algorithm (e.g. \cite{waterfill}) cannot be used to obtain the power allocation vector. Further, the rate constraint in \eqref{maxeq} also has a non-linear product involved and an expression for the power limited by the constraint cannot be directly obtained. Note that direct application of the CWF algorithm with the mean value of the channel fading coefficients will prove to be  sub-optimal (as we will see in section \ref{StatanalSimRes}). Our key contribution in this paper is that we develop a rational approximation to the non-linear term $\exp\left(\frac{k_m^{rd}}{P_m}\right)E_1\left(\frac{k_m^{rd}}{P_{m}}\right)$, so that we can solve the optimization problem in \eqref{maxeq}. Specifically we approximate the non-linear product $\exp\left(\frac{k_m^{rd}}{P_{m}}\right)E_1\left(\frac{k_m^{rd}}{P_{m}}\right)$ in \eqref{maxeq} by using a rational function\footnote{In the above approximation, we have limited the degree of the rational function to 1, as any higher degree rational function leads to a polynomial equation of degree five or higher when we try to solve the Lagrange's function (discussed later). This leads to an intractable solution for the power allocation scheme at the relay.} of the form,
\begin{equation}\label{approx}
\exp\left(\frac{k_{m}^{rd}}{P_{m}} \right)E_1\left ( \frac{k_{m}^{rd}}{P_{m}} \right ) =  \left(\frac{a_{m}\left(\frac{k_{m}^{rd}}{P_{m}}\right)+ b_{m}}{c_{m}+\left(\frac{k_{m}^{rd}}{P_{m}}\right)}\right)+ \epsilon,
\end{equation}
where $a_{m}, b_{m}$, $c_{m}$ are constants, and $\epsilon$ is the error in approximation. The approximation in \eqref{approx} is based on minimizing the error  $\epsilon$. As a measure of the approximation of the estimation of $a_m$, $b_m$ and $c_m$, we computed the root mean squared error (RMSE) on the approximation as, $\text{RMSE} = \sqrt{ \frac{[S(a_m, b_m, c_m)]^2}{n}},$ where $S(a_m, b_m, c_m)$  and $n$ are defined in Appendix~\ref{appendx1}. For the approximation in \eqref{approx}, the RMSE was found to be  $ < 10^{-3}$ (when the SNR at the destination in the range of $-15$ to $30$ dB) leading to error in approximation $\epsilon < 10^{-3}$. The difference between the achievables rates with the approximation and with the exponential integral function is therefore $<10^{-3}$.
Note that in \eqref{approx}, the constants $a_{m}, b_{m}$ and $c_{m}$ depend on the ratio $\frac{k_{m}^{rd}}{P_{m}}$ (denoted by $\Delta = \frac{k_{m}^{rd}}{P_{m}}$). Note also, that for different values of $\Delta$ we may need to find different values of the constants $a_{m}, b_{m}$, $c_{m}$, which minimizes $\epsilon$. We start by evaluating an estimate of the power (denoted as $P_m^{est}$) allocated to the $m$th user. $P_m^{est}$ is found by setting $h_m^{sr}$, $h_m^{sd}$ and $h_m^{rd}$ to their mean value in \eqref{achRate2}, and by using a CWF algorithm. Note that $P_m^{est}$ is only used to obtain the constants $a_{m}, b_{m}$, $c_{m}$ so that the approximation in \eqref{approx} can be used. The constants $a_{m}, b_{m}$ and $c_{m}$, for different ranges of $\Delta$ are then found by using the lookup Table~\ref{StatTable1}. Appendix~\ref{appendx1}, describes the procedure to obtain Table~\ref{StatTable1}. The algorithm for determining $a_m, b_m$ and $c_m$ is described below.
\begin{algorithm}
\caption{Algorithm for determining $a_m, b_m$ and $c_m$.}
\label{ambmcmAlg}
\textbf{Step 1:} Set $h_m^{sr}= h_m^{sd} = h_m^{rd} = \frac{\pi}{2\sqrt 2}$ for all $m \in \{1,...,M\}$. Note that $\Delta_1$, $\Delta_2$, $\Delta_3$ and $\Omega_m$ are temporary variables used in the algorithm.

\textbf{Step 2:} Obtain $P_m^{est}$ using the CWF algorithm, under the power constraint $\sum_{m=1}^{M}{P_m^{est}} = P_{r}$.

\textbf{Step 3:} Compute $\Omega_m = 10\log_{10}\left(\frac{k_{m}^{rd}}{P_m^{est}} \right)$, for all $m \in \{1,...,M\}$. For all $m \in \{1,...,M\}$ do the following. If $\Omega_m \in \Delta_1$, set $a_m = a(\Delta_1) , b_m = b(\Delta_1)$ and $c_m = c(\Delta_1)$. Else, if $\Omega_m \in \Delta_2$, set $a_m = a(\Delta_2) , b_m = b(\Delta_2)$ and $c_m = c(\Delta_2)$. Otherwise set $a_m = a(\Delta_3) , b_m = b(\Delta_3)$ and $c_m = c(\Delta_3)$.
\end{algorithm}
Note that Table.~\ref{StatTable1} is pre-computed and stored in the memory of the relay.

\section{Lagrangian-Based Power Allocation Scheme (PAS-1)}\label{PAS-1}
We now develop a Lagrangian-based power allocation scheme at the relay using the approximation given in \eqref{approx}. To find the power allocation scheme which maximizes the achievable rate $R_{2m}^{+}$ in \eqref{maxeq}, we set up the generalized Lagrange's multiplier function for non-linear optimization as follows,
\begin{multline}\label{lagrange}
L(P,\mu,\nu, \tau)= -\sum_{m=1}^{M} \log(e)\left [ \frac{\exp\left ( \frac{k_{m}^{rd}}{P_{m}}\right )E_{1}\left ( \frac{k_{m}^{rd}}{P_{m}} \right ) - \exp\left ( \frac{k_{m}^{sd}}{P_{s}}\right )E_{1}\left ( \frac{k_{m}^{sd}}{P_{s}} \right )}{\left(1- \frac{P_{s}k_{m}^{rd}}{P_{m}k_{m}^{sd}}\right)} \right ]\\
+ \sum_{m=1}^{M}\mu_{m}(-P_{m})
-\sum_{m=1}^{M}\nu_{m}\left(R_{2m}^{+} - R_{1m} + R_{2m}^{-}\right)+ \left[\left(\sum_{m=1}^{M}\tau P_{m} \right )-P_r \right] ,
\end{multline}
with definitions $P = [P_{1},...,P_{M}]$, $\mu  = [\mu_{1},...,\mu_{M}]$ and $\nu  = [\nu_{1},...,\nu_{M}]$, where $\mu$, $\nu$ and $\tau$ are Lagrange multipliers associated with the constraints in \eqref{maxeq}.
We obtain the necessary and sufficient Karush-Kuhn-Tucker (KKT) conditions as,
\begin{align}
\left.\begin{matrix}\label{KKT1}
\{\frac{\partial L(P,\mu,\nu,\tau)}{\partial P_{m}}, \mu_{m}P_{m}\} = 0\\
\{-\mu_{m}, -\nu_{m} \}\leq 0\\
\nu_{m}\left(R_{2m}^{+} - R_{1m} + R_{2m}^{-}\right)\leq 0\\
\end{matrix}\right\} m = 1,...,M.
\end{align}
To solve the Lagrangian function in \eqref{lagrange}, we use the approximation in \eqref{approx}.
Let $\phi_m(d_m^{rd}, d_m^{sd}, P_{s}) = \{[0, P_r]:\phi_m(\cdot) \in \mathbb{R}\}$ be a function defined in Appendix~\ref{appendix2}, which denotes the power allocated to the $m$th user after finding the optimal Lagrange's multiplier $\tau^*$, that satisfies the constraint $\sum_{m=1}^{M}{P_{m}} = P_{r}$. Similarly, let $\pi_m^c(d_m^{rd}, d_m^{sr}, d_m^{sd}, P_{s})= \{[0, P_r]: \pi_m^c(\cdot) \in \mathbb{R}\}$ be another function defined in Appendix~~\ref{appendix2}, which denotes the power obtained by using substituting the approximation in \eqref{approx} into the rate constraint in \eqref{maxeq}, and solving for $P_m$. We propose the following theorem.

\emph{Theorem 1}: The power allocation scheme at the relay that approximately obtains the maximum throughput of an $M$-1-1 system with statistical-CSI is given by,
\begin{equation}\label{apprxpowth}
P_{m} = \begin{cases}
&\phi_m(d_m^{rd}, d_m^{sd}, P_{s}),\;\;\;\;\;\;\;\;\;\;\;\text{if}\;\;\phi_m(d_m^{rd}, d_m^{sd}, P_{s}) < \pi_m^c(d_m^{rd}, d_m^{sr}, d_m^{sd}, P_{s})\\
&\pi_m^c(d_m^{rd}, d_m^{sr}, d_m^{sd}, P_{s}),\;\;\;\;\;\text{if}\;\phi_m(d_m^{rd}, d_m^{sd}, P_{s})> \pi_m^c(d_m^{rd}, d_m^{sr}, d_m^{sd}, P_{s}).\\
\end{cases}
\end{equation}

\emph{Proof:} See Appendix~\ref{appendix2}.

The PAS-1 algorithm is described below.
\begin{algorithm}[h]
\caption{PAS-1 Algorithm.}
\label{lpasAlg}
\textbf{Step 1:} Initialize $P_{rem} = P_r$ and  $M^* = \{1,...,M \}$. Note that $P_{rem}$, $M^*$, $P_{ext}$ are variables which are function of the iterations between the steps.

\textbf{Step 2:} Obtain $a_m, b_m$ and $c_m$ using Alg.~\ref{ambmcmAlg} for $m \in M^*$. Compute $\pi_m^c(\cdot)$.

\textbf{Step 3:} Use bisection search method to compute $P_m = \phi_m(\cdot)$ for $m \in M^*$ subject to $\sum_{m\in M^*}{P_{m}} = P_{rem}$.

\textbf{Step 4:} Find the set $\mathcal{M}$ of users, which have $\phi_m(\cdot) > \pi_m^c(\cdot)$. If the number of elements in $\mathcal{M}$ is equal to either $0$ or $M$, then exit.

\textbf{Step 5:} Let $P_{m} = \pi_m^c(\cdot)$, for $m \in \mathcal{M}$. Calculate the extra power $P_{ext} = \sum_{m\in\mathcal{M}}\left[\phi_m(d_m^{rd}, d_m^{sd}, P_{s})-\pi_m^c(\cdot)\right]$.

\textbf{Step 6:} Obtain the set of channels $\bar{\mathcal{M}}$ complementary to $\mathcal{M}$. For the set $\bar{\mathcal{M}}$ compute the total power $P_{\bar{\mathcal{M}}}$  as $P_{\bar{\mathcal{M}}}~=~\sum_{m\in\bar{\mathcal{M}}}{P_{m}}$. Compute $P_{rem}~=~P_{ext} +P_{\bar{\mathcal{M}}}$ and set $M^* = \bar{\mathcal{M}}$. Goto Step 2.
\end{algorithm}
Even though PAS-1 provides a system achievable rate approximately equal to $R_{PAS-0}$, as we will see in section \ref{StatanalSimRes}, the computations required in the PAS-1 algorithm may still be high due to the iterative bisection search in Step 2. The system achievable rate is insensitive to exact power allocation for high values of received SNR at the destination since the achievable rate is a logarithmic function of the power. This motivates us to investigate a lower computational power allocation scheme that can perform close to PAS-1 in the following section.

\section{Low-Computational Power Allocation Scheme  (PAS-2)}\label{nopas}

Using the results from PAS-1, we now develop another power allocation scheme (PAS-2) at the relay which achieves a system achievable rate within $5-10\%$ of the achievable rate $R_{PAS-0}$, but with significantly lower computations. To develop PAS-2, we build on the work proposed in \cite{WeiYu}, in which a low-complexity power allocation scheme was proposed for a single transmitter and receiver system with multicarrier modulation and Intersymbol Interference (ISI) channels. In \cite{WeiYu}, the transmitter allocates zero power to subchannels with channel gains greater than a threshold, and equal power to the remaining subchannels. This concept is motivated by the fact that $R$ is insensitive to exact power allocation for high values of SNR. However, the system model considered in \cite{WeiYu} is different compared to our system, and as such the power allocation scheme cannot be directly applied to our $M$-1-1 system. Any power allocation scheme in our $M$-1-1 system needs to take into account the rate constraint in \eqref{maxeq}. Note that the combined channel gain between the $m$th source and $d$; and the channel between $r$ and $d$ can be obtained by rearranging the second term of \eqref{achRate2} and is given by,
\begin{equation}\label{G_m}
G_{m} = \frac{(d_{m}^{sd})^\alpha\left | h_{m}^{rd} \right |^2}{(d_{m}^{rd})^\alpha\left [ P_{s} \left | h_{m}^{sd} \right |^2 + N_{d}(d_{m}^{sd})^\alpha\right ]}.
\end{equation}
Our power allocation scheme 2 (PAS-2) can be outlined as follows. We compute $G_m$ in \eqref{G_m} using the mean value of $h_m^{sr}$, $h_m^{sd}$ and $h_m^{rd}$. We then sort the users based on their channel gains ($G_m$) and find the set of users who should be allocated non-zero power. We divide the power equally among the set of users, who must receive non-zero power. We then find the set of users $\mathcal{M}$, whose allocated power exceeds the power limited by the rate constraint function $\pi_m^c(\cdot)$. The remaining power after applying the constraints is computed and redistributed equally among the users in the complementary set $\bar{\mathcal{M}}$. This is done iteratively until all available power is distributed. The PAS-2 algorithm is described below.
\begin{algorithm}[h]
\caption{PAS-2 Algorithm.}
\label{nopasAlg}
\textbf{Step 1:} Initialize $P_m = 0$, $P_{rem} = P_r$ and  $M^* = \{1,...,M \}$. Note that $P_{rem}$, $M^*$, $P_{ext}$ and $P_{est}$ are variables which are function of the iterations between the steps.\\
\textbf{Step 2:} Compute $G_m$ for $m \in\{1,...,M\}$, using \eqref{G_m} with $h_m^{sr}$, $h_m^{sd}$ and $h_m^{rd}$ set to their mean value. Sort the channel gains, such that $G_1 \ge G_2 \ge ...\ge G_m$ for $m \in M^*$.\\
\textbf{Step 3:} Set $P_m = P_{rem}/|M^*|$, where $|M^*|$, denotes the cardinality of set $M^*$.

\textbf{Step 4:} If ($1/G_{|M^*|} \ge P_r + 1/G_{1}$), then $|M^*| = |M^*|-1$, Goto Step 3.

\textbf{Step 5:} Set $P_{est} = P_{m}$. Obtain $a_m, b_m$ and $c_m$ from Step 3 of Alg.~\ref{ambmcmAlg}. Compute $\pi_m^c(\cdot)$.

\textbf{Step 6:} Set $M^* = M^*+1$. Find the set $\mathcal{M}$ of the channels, which have $P_m > \pi_m^c(\cdot)$ for $m \in \mathcal{M}$. If the number of elements in $\mathcal{M}$ is equal to either $0$ or $M$, then exit.

\textbf{Step 7:} Let $P_{m} = \pi_m^c(\cdot)$, for $m \in \mathcal{M}$ and calculate the extra power as $P_{ext}~=~\sum_{m\in\mathcal{M}}\left[P_m -\pi_m^c(\cdot)\right]$.

\textbf{Step 8:} Obtain the set of users $\bar{\mathcal{M}}$ complementary to $\mathcal{M}$. Compute the total power in set $\bar{\mathcal{M}}$ as  $P_{\bar{\mathcal{M}}}~=~\sum_{m\in\bar{\mathcal{M}}}{P_{m}}$. Compute the remaining power $P_{rem} = P_{ext} + P_{\bar{\mathcal{M}}}$ and set $M^* = \bar{\mathcal{M}}$. Goto Step 2.
\end{algorithm}

\section{Computation Complexity}\label{complexity}

\subsection{Computation Complexity of PAS-0 (Optimal)}

The exact computational complexity of any numerical method of optimization is difficult to obtain as it depends on the number of times that the objective function and its derivatives are computed. It also depends on how many iterations are required to reach some stopping/convergence criterion and how many constraints are active during an iteration. In general this will be of the order $\mathcal O(M^3)$ (e.g. interior point method \cite{optbook}).

\subsection{Computation Complexity of PAS-1}

The complexity of the PAS-1 algorithm is mainly in Step 2 and Step 3. Obtaining $a_m$, $b_m$ and $c_m$ in Step 2 involves the CWF algorithm, whose complexity is of order $\mathcal O(M\log M)$ \cite{WFComplexity}. In Step 2 of PAS-1 algorithm, the optimum value of Lagrangian multiplier $\tau^*$ should be searched to compute $\phi_m(d_m^{rd}, d_m^{sd}, P_{s})$, such that the constraint  $\sum_{m = 1}^{M}\phi_m(d_m^{rd}, d_m^{sd}, P_{s}) = P_{r}$ is satisfied. We used a bisection search method~\cite{bisection} to find $\tau^*$. The complexity of an efficient bisection search algorithm is of the order $\mathcal O(M \log M)$~\cite{Krongold}. The total complexity of the PAS-1 algorithm is then of the order $\mathcal O(KM + 2KM \log M)$, where $K$ ($K < M$) is the number of iterations required between Step 2 and 7 in the PAS-1 algorithm. Through our simulations (discussed in section~\ref{StatanalSimRes}), we found that the maximum value of $K$ is $K = 20$ for $M = 100$ users.

\subsection{Computation Complexity of PAS-2}

The PAS-2 algorithm does not require the bisection search of $\tau^*$ as in the PAS-1 algorithm. The complexity of the sorting of users in Step 1 is $\mathcal O(M\log M)$. The other complexity is in step 3 and 4 being iteratively executed. Step 3 and 4 has a complexity of $\mathcal O(M\log M$)~\cite{WeiYu}. The total computational complexity of the algorithm is then of the order $\mathcal O(KM~+~2KM\log M)$, where $K$ ($K < M$) is the number of iterations required between Step 2 and 7 in the PAS-2 algorithm. Here, we will show that the PAS-2 algorithm requires far less computations then the PAS-1 algorithm. This is because in Step 3 of the PAS-1 algorithm, we need to compute the rate $R_{2m}^{+}$ for each step of bisection search to find the new value of $\tau$ (see \cite{Krongold}). However, in the PAS-2 algorithm the optimal water-level is found without actually computing $R_{2m}^{+}$ in each step, and is therefore free of logarithmic operations. Further, during each iteration, the PAS-2 algorithm is free of logarithm and cube root operations, and requires only 1 square root computation per user compared to 8 square roots computations per users in the in Step 2 of the PAS-1 algorithm. The number of computations required per iteration for the PAS-1 and PAS-2 algorithms are summarized in Table.~\ref{StatTable2}. We can see in Table.~\ref{StatTable2} that the PAS-2 algorithm requires significantly lower number of computations compared to the PAS-1 algorithm.
We also measured the execution times required for each of the power allocation algorithms in MATLAB. The results plotted in Fig.~\ref{exectimeMatlab} show that PAS-0 takes over 1000x more time than PAS-1 and PAS-2 algorithms making them suitable for practical implementations.

\section{Numerical Results}\label{StatanalSimRes}

Here we present the analytical and simulation results for the system achievable rate using the power allocation schemes developed in the previous sections. The various parameters were configured as follows. The path-loss exponent was set to $\alpha = 2$, $P_{s}$ was set to 5 and $N_{d}$ and $N_{r}$ was set to 1. Note that, we have investigated the achievable rates for various values of the above parameters. For convenience, we make the following notations. We denote the system achievable rate obtained with PAS-0 (numerical optimization), PAS-1 and PAS-2 as $R_{PAS-0}$, $R_{PAS-1}$ and $R_{PAS-2}$ respectively. We also implemented a power allocation scheme as described in \cite{WeiYu} with channel fading coefficients set to their mean value. This corresponding system achievable rate will be denoted as $R_{SUBOP}$. Note that for PAS-0, we used the Interior-Point Algorithm \cite{optbook} to obtain $R_{PAS-0}$.

\begin{figure}[t]
\par
\begin{center}
{\includegraphics[width=0.5\textwidth]{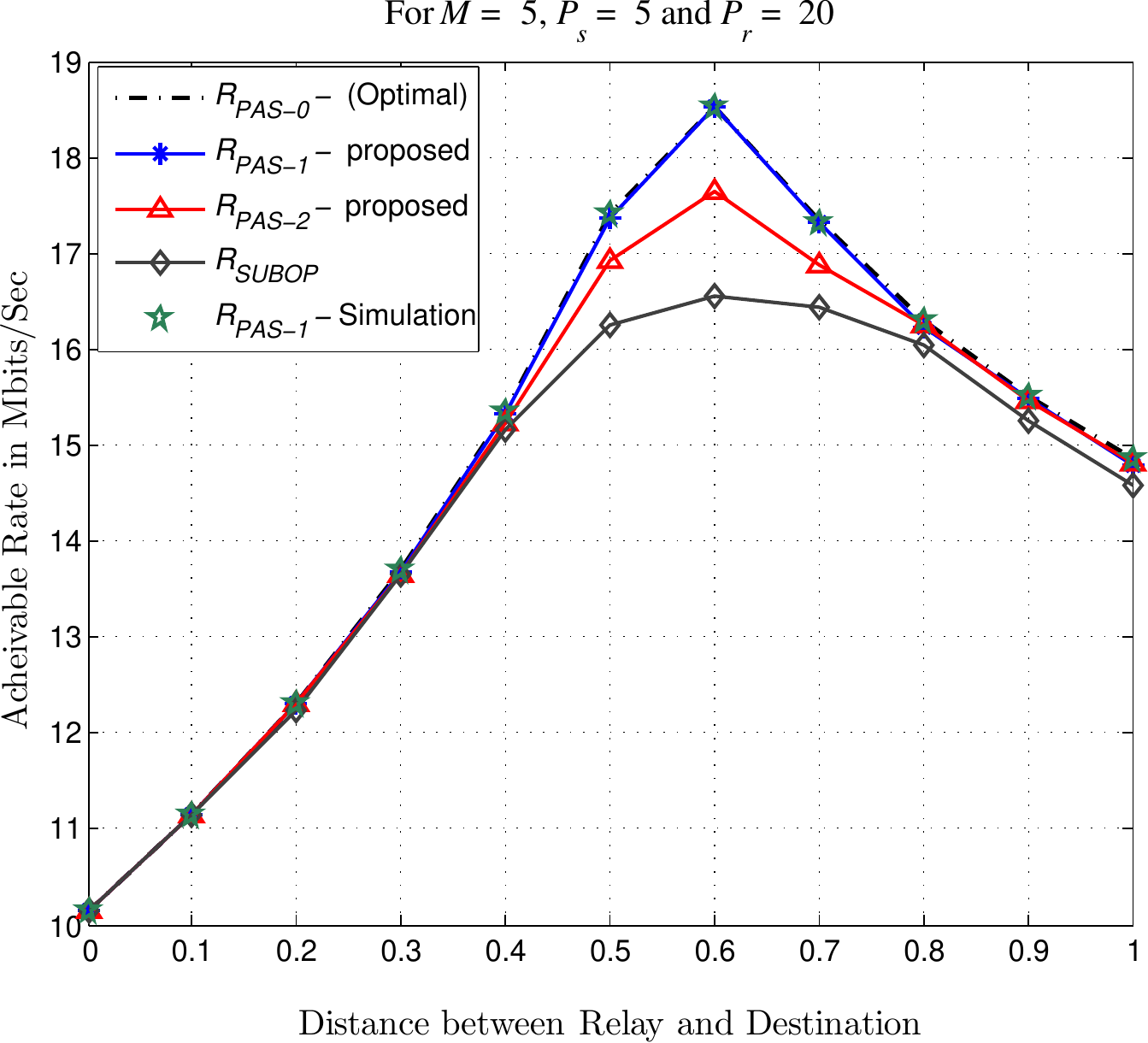}}
\end{center}
\caption{Achievable Rates with different power allocation schemes for $M =5$ users, $P_s =5$ and $P_r = 20$.}
\label{statResM5}
\end{figure}

\begin{figure}[p]
\centering


\begin{tabular}{cc}
\includegraphics[width=0.5\textwidth]{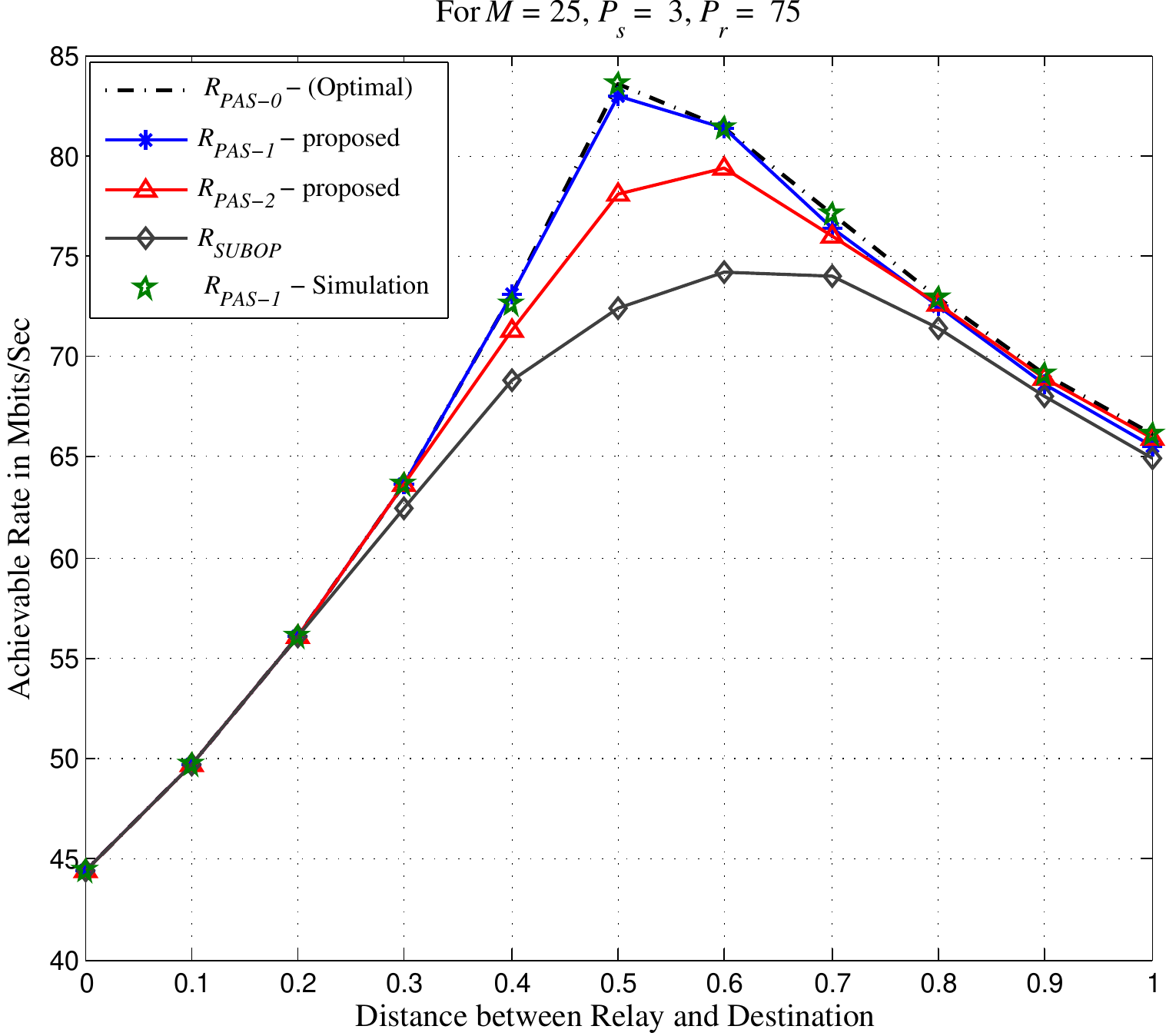}
\end{tabular}
\caption{Achievable Rates with different power allocation schemes for $M =25$ users, $P_s =3$ and $P_r = 75$.}
\label{statResM25}
\end{figure}

\begin{figure}[p]
\centering

\begin{tabular}{cc}
\includegraphics[width=0.5\textwidth]{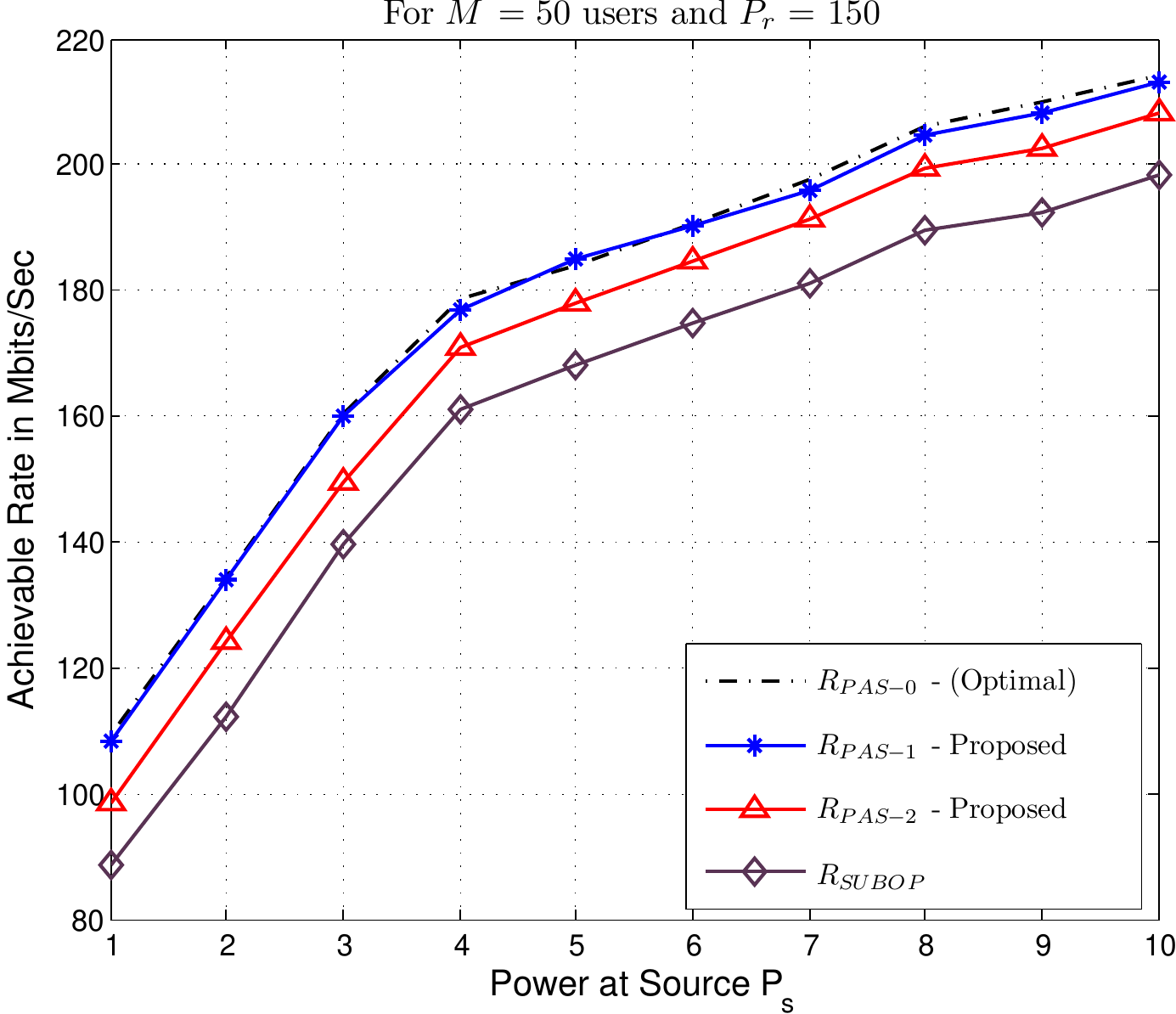}
\end{tabular}
\caption{Achievable Rates with different power allocation schemes plotted against $P_s$ for $M =50$ users and $P_r = 200$.}
\label{statResRvsPs}

\begin{tabular}{cc}
\includegraphics[width=0.5\textwidth]{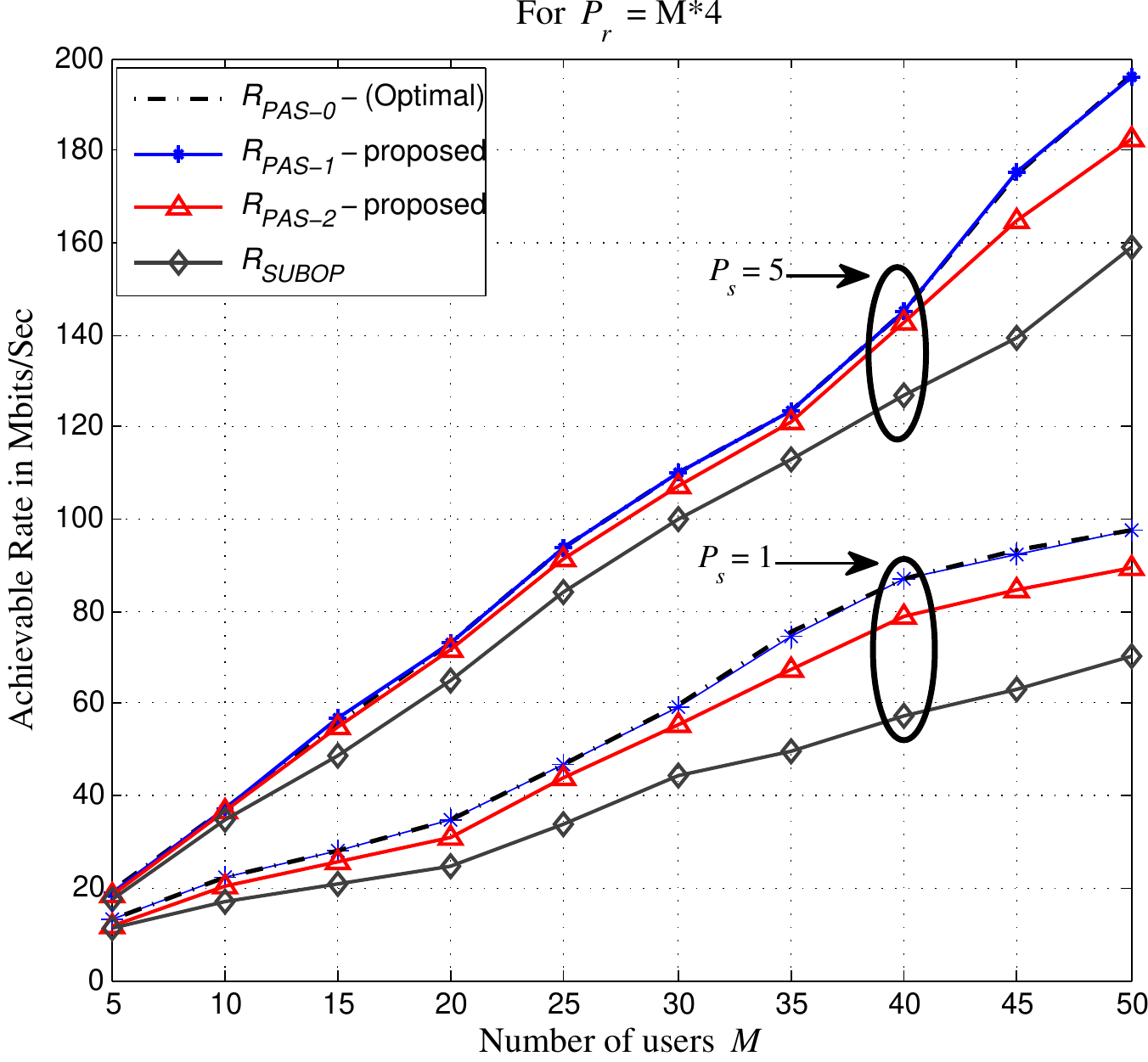}
\end{tabular}
\caption{Achievable Rates with different power allocation schemes for different $M$ with $P_r = 4M$, plotted for $P_s = 1$ and $P_s = 5$.}
\label{statResFixedRelay}
\end{figure}

We also performed Monte Carlo simulations to verify our analysis. In our simulation setup, the normalized distance between relay and destination is initially set to 1 and $M$ sources are randomly distributed, within a circle of radius 0.5 centered around the relay. The relay is then moved along the straight line towards the destination. For each position of the relay, the system achievable rate is computed as follows. The channel coefficients $h_{m}^{sr}$, $h_{m}^{sd}$ and $h_{m}^{rd}$ are drawn from a Rayleigh distribution. For each channel realization, $R_{m}$ for $m = 1$  to $M$ users is computed from \eqref{achRate2} using the PAS-1 algorithm. $R$ was obtained using \eqref{R_summation} and averaged over 2000 channel realizations. The users were redistributed and the simulations repeated over 2000 trails. The results are plotted in Fig.~\ref{statResM5} for $M =5$ users and $P_r = 20$.  $R_{PAS-0}$, $R_{PAS-1}$, $R_{PAS-2}$ and $R_{SUBOP}$ for $M=25$, $P_s =3 $ and $P_r =75$ are plotted in Fig.~\ref{statResM25}. $R_{PAS-0}$, $R_{PAS-1}$, $R_{PAS-2}$ for different values of $P_s$ is plotted in Fig.~\ref{statResRvsPs}, with $M = 50$ and $P_r = 200$. Note that we have assumed a subchannel bandwidth of 1MHz (Mega Hertz) and that the achievable rate is in Mbits/sec. Note also that with increasing values of $M$,  $R_{SUBOP}$ is significantly lower than $R_{PAS-0}$, whereas $R_{PAS-1}$ is almost equal to $R_{PAS-0}$.

\begin{figure}[p]
\centering

\begin{tabular}{cc}
\includegraphics[width=0.5\textwidth]{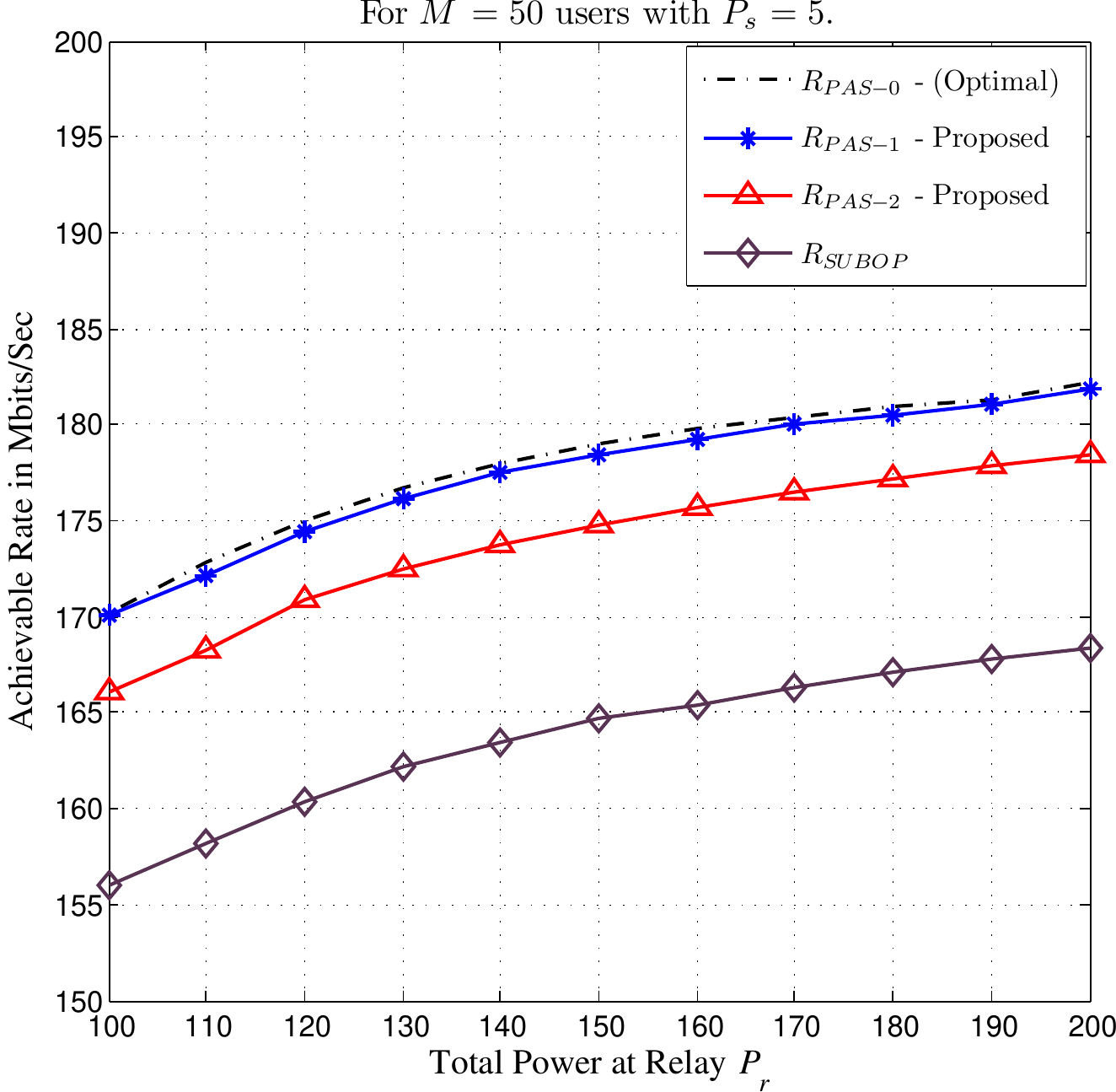}
\end{tabular}
\caption{Achievable Rates with different power allocation schemes plotted against $P_r$ for $M =50$ users and $P_s = 5$.}
\label{statResRvsPr}

\begin{tabular}{cc}
\includegraphics[width=0.5\textwidth]{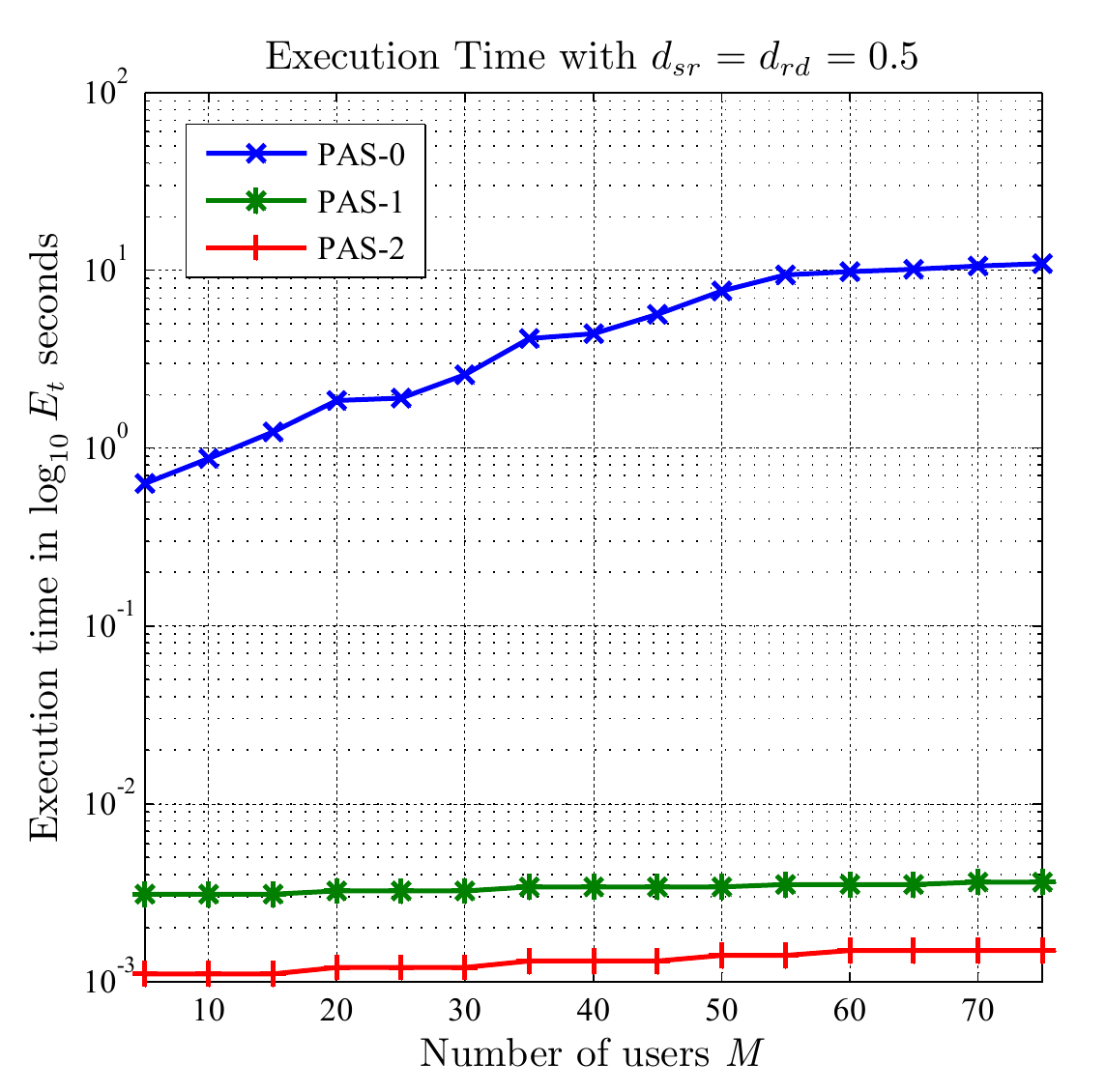}
\end{tabular}
\caption{MATLAB execution times with various power allocation schemes.}
\label{exectimeMatlab}

\end{figure}

The system achievable rate as function of $M$ is plotted in Fig.~\ref{statResFixedRelay} with relay placed midway on the line between the center of the source circle and the destination. Note that with PAS-1, the system achievable rate $R_{PAS-1}$ in all the results is approximately equal to $R_{PAS-0}$. $R_{PAS-2}$ obtained with PAS-2 is within $\approx 5\%$ of $R_{PAS-0}$. Note that the rates plotted in all the figures is the sum of the achievable rates for $M$ users. The transmit powers at the source and relay and normalized with a reference power of 1 mW. Similiarly the noise powers at the relay and destination are also normalized with a reference power of 1mW. We also plotted the results for various values of $M$, with the relay placed midway between the source circle and destination, in Fig.~\ref{statResFixedRelay}. Note in Fig.~\ref{statResFixedRelay}, $R_{PAS-1}$ is still approximately equal to $R_{PAS-0}$, whereas $R_{PAS-2}$ is still within $5-10\%$ of $R_{PAS-0}$. Also note that $R_{SUBOP}$ proves to be sub-optimal. The loss in system throughput with this sub-optimal power allocation is up to $ \approx 30\%$ relative to using PAS-1, for $M= 50$ users in Fig.~\ref{statResFixedRelay}. We anticipate this loss to be higher for higher values of $M$. Using the PAS-1 or the PAS-2 algorithm, then turns out to be a tradeoff between low computations and system achievable rate. The system achievable rate with different power allocation schemes for $P_s =5$, $M = 50$ users for different values of  $P_r$ is shown in Fig.~\ref{statResRvsPr}. We can see from Fig.~\ref{statResRvsPr} that the PAS-1 algorithm delivers a throughput close to $R_{PAS-0}$ and that the PAS-2 algorithm is within $5-10\%$ of $R_{PAS-0}$. Therefore, the PAS-1 algorithm is suitable in an $M$-1-1 system, where the priority is to maximize system throughput, whereas the PAS-2 algorithm is suitable for an $M$-1-1 system where lower computations in power allocation scheme are essential.

\section{Conclusions}\label{StatConcls}

In this paper, we considered a multi source decode-and-forward, full-duplex relay system (M-1-1 system) with statistical-CSI. We investigated the achievable rate, $R$, of an $M$-1-1 system with statistical-CSI, where all channels undergo independent Rayleigh fading. We showed how $R$ can only be maximized using numerical power allocation schemes which has a high-complexity of order $\mathcal O(M^3)$. We introduced a rational approximation in the achievable rate analysis, based on which we developed two low-complexity power allocation schemes at the relay that obtain a system throughput close to the maximum ($R_{PAS-0}$). Specifically, we developed a Lagrangian-based power allocation scheme (PAS-1), which obtains a system achievable rate approximately equal to $R_{PAS-0}$ for all practical purposes. Utilizing the results derived in PAS-1, we developed another power allocation scheme (PAS-2), which delivers a system achievable rate within $5-10\%$ of $R_{PAS-0}$, but with a significantly lower number of computations.

Most importantly we showed that the complexity of the PAS-1 and PAS-2 algorithms is of order $\mathcal O(M\log M)$. We provided simulations results to justify our analysis. We showed how PAS-1 is suitable in an $M$-1-1 system with priority on system throughput, whereas PAS-2 is suitable for an $M$-1-1 system where lower computations in power allocation scheme are essential. The power allocation schemes, PAS-1 and PAS-2, developed in this paper will be of value to design and implementation of an real-time multi-user relay systems operating under realistic channel conditions.

\section*{Acknowledgments}

The authors would like to thank the anonymous reviewers whose input and suggestions greatly improved the quality of this paper. The authors would also like to thank the University of New South Wales and the Australian Research Council (ARC Grant number DP0879401) for supporting this research work.

\appendices

\section{Generating Table~\ref{StatTable1}}\label{appendx1}
We approximated the non-linear product $\exp\left(\frac{k_m^{rd}}{P_m}\right)E_1\left(\frac{k_m^{rd}}{P_{m}}\right)$ in \eqref{maxeq} by a curve fitting technique. The average received SNR (in dB) at the receiver (destination) is given as, $\gamma_{SNR} = P_{t} - P_{L} + G_{t} + G_{r} - N_d$, where $P_{t} = 10\log_{10}(P_m)$, $P_L = \alpha \log_{10}(d_m^{rd})$ (in dB), and $G_t$ and $G_r$ are transmit and receive antenna gains (in dB), respectively. In the ratio $\frac{k_{m}^{rd}}{P_{m}} = \frac{(d_m^{rd})^\alpha N_d}{P_m}$, $(d_m^{rd})^\alpha$ represents the path-loss between the $r$ and $d$. Therefore, the ratio $\frac{k_{m}^{rd}}{P_{m}}$ scales as $1/\gamma_{SNR}$. Since $\gamma_{SNR}$ is typically in the range of -15 to 30dB, we are interested in finding the constants $a_{m}, b_{m}$ and $c_{m}$, when $\frac{k_{m}^{rd}}{P_{m}}$ is in the range of $-15$ to $30$ dB. However, it is not possible to find one set of values for $a_{m}, b_{m}$ and $c_{m}$ with an acceptable error in approximation (i.e., $\epsilon \le 10^{-3}$) over the entire range of $\frac{k_{m}^{rd}}{P_{m}}$. We therefore divide $\frac{k_{m}^{rd}}{P_{m}}$ into 3 ranges as $\{\Delta_1~\in~\frac{k_{m}^{rd}}{P_{m}}~:~-15dB <\Delta_{1} < 0dB\}$, $\{\Delta_{2} \in \frac{k_{m}^{rd}}{P_{m}} : 0dB < \Delta_2 < 15dB\}$ and $\{\Delta_3~\in~\frac{k_{m}^{rd}}{P_{m}}~:~15dB~<\Delta_3~<~30dB\}$. Note that splitting $\frac{k_{m}^{rd}}{P_{m}}$ into more than 3 ranges will although increase the precision of approximation in \eqref{approx}, but would not alter the final results significantly. This is because the error in approximation $\epsilon$  over all three ranges  is already less than $10^{-3}$ and increasing the number of ranges k, will although reduce $\epsilon$, but will not affect $R$ significantly. The values of $a_{m}, b_{m}$ and $c_{m}$ are found as follows. 
Let us define, $\Delta_k^i$ as a discrete value in the range $\Delta_k$. We perform a non-linear least squares analysis by minimizing the sum of non-linear least squares defined as,
\begin{equation}\label{nonlinearsq}
 S(a, b, c) =  \sum_{i=1}^{n}\left [ \exp\left(\Delta_k^i\right)E_1\left ( \Delta_k^i \right ) - \left(\frac{a\Delta_k^i + b}{c+\Delta_k^i}\right)\right ]^2,
\end{equation}
where $n$ is the number of discrete values in  $\Delta_k = [\Delta_{k}^{1},\Delta_{k}^{2},...,\Delta_{k}^{n-1}, \Delta_{k}^{n}]$. We used the Levenberg-Marquardt algorithm to minimize the sum of non-linear squares $S(a, b, c)$ in \ref{nonlinearsq}, by setting $n = 10^4$. Table.~\ref{StatTable1} lists the values of $a, b$, $c$ for different ranges of $\Delta_k$ with $\epsilon \le \times10^{-3}$. We stress here the fact that the values of $a, b$ and $c$ are pre-computed and stored in the memory of the relay. Note that if $h_{sr}$, $h_{sd}$ and $h_{rd}$ are Rician distributed with parameter $\nu$, then $|h_{sr}|^2, |h_{sd}|^2$, and $|h_{rd}|^2$ follow non-central $\chi^2$ distribution with two degrees of freedom and non-centrality parameter $\nu^2$. To find the average achievable rate of the whole system, the channel rates have to be integrated over all channel states, which does not yield a closed form expression. Therefore, the proposed approximation cannot be extended to a channel with Rician distribution.

\section{Proof for Power Allocation Theorem 1}\label{appendix2}

Using the approximation in \eqref{approx} in the partial derivative of the Lagrangian function with respect to $P_m$ in \eqref{KKT1}, leads to
\begin{equation}\label{diffRes}
\mu_m = (\nu_m+\tau) - \frac{(\log e) k_m^{rd}k_m^{sd} \left(\left(\frac{a_{m}k_{m}^{rd}+ b_{m}P_{m}}{c_{m}P_{m}+k_{m}^{rd}}\right)-\beta\right)}{\left(P_{m} -k_m^{rd}k_m^{sd}  \right)^2 }
+\frac{\log e \left(P_{m}-k_m^{rd}\left(\frac{a_{m}k_{m}^{rd}+ b_{m}P_{m}}{c_{m}P_{m}+k_{m}^{rd}}\right)\right)}{P_{m}(P_{m}-k_m^{rd}k_m^{sd})},
\end{equation}
where $\beta = \exp\left ( \frac{k_{m}^{sd}}{P_{s}}\right )E_{1}\left ( \frac{k_{m}^{sd}}{P_{s}}\right)$. The condition $\mu_mP_{m} = 0$ leads to either $P_{m} = 0$ or $\mu_m = 0$. Setting  $\mu_m = 0$  in \eqref{diffRes} and after some algebra we get,
\begin{multline}\label{muzero}
P_{m}^4(\tau+\nu_m)c_{m}+ P_{m}^3[k_{m}^{rd}(\tau+\nu_m)-2c_{m}k_{m}^{rd}k_{m}^{sd}(\tau+\nu_m) -c_{m}\log e]
+ P_{m}^2\left[ c_{m}(k_{m}^{rd}k_{m}^{sd})^2(\tau+\nu_m) - 2(k_{m}^{rd})^2k_{m}^{sd}(\tau+\nu_m)\right.\;\;\;\;\;\;\;\;\;\;\;\;\;\;\;\;\;\;\;\;\;\;\;\;\;\;\;\;\;\;\;\;\;\;\;\;\\
\left.+ c_{m}k_{m}^{rd}k_{m}^{sd}(1-\beta)\log e + b_{m}k_{m}^{rd}k_{m}^{sd}\log e +b_{m}k_{m}^{rd}\log e (b_{m}-1)\right] \\ + P_{m}[(\tau+\nu_m)(k_{m}^{rd})^3(k_{m}^{sd})^2 + a_{m}(k_{m}^{rd})^2(k_{m}^{sd} +1)\log e -k_{m}^{rd})^2k_{m}^{sd}(b_{m}- \beta+1)\log e ]\;\;\;\;\\
-a_{m}(k_{m}^{rd})^3k_{m}^{sd}=0.\;\;\;\;\;\;\;\;\;\;\;\;\;\;\;\;\;\;\;\;\;\;\;\;\;\;\;\;\;\;\;\;\;\;\;\;\;\;\;\;\;\;\;\;\;\;\;\;\;\;\;\;\;
\end{multline}

The rate constraint in \eqref{KKT1} leads to two cases, 1) $\nu_{m} = 0$ or
2) $\left(R_{2m}^{+} - R_{1m} + R_{2m}^{-}\right) = 0$. The second case leads to,
\begin{equation}\label{quadroot}
P_{m} = \pi_m^c(k_m^{rd}, k_m^{sr}, k_m^{sd}, P_{s}) = \frac{k_{2}-\sqrt{k_{2}^{2}+4k_{1}k_{3}}}{2k_1}.
\end{equation}
where $k_{1} = b_{m}-c_{m}\psi$, $k_{2} =  k_{m}^{rd}(a_{m}-\psi)+ \frac{c_{m}P_{s}k_{m}^{rd}}{k_{m}^{sd}}(\beta-\psi)$, and $k_{3} = \frac{P_{s}(k_{m}^{rd})^2}{k_{m}^{sd}}\left(\psi-\beta\right)$,
and where, $\psi = \exp\left ( \frac{k_{m}^{sr}}{P_{s}}\right )E_{1}\left ( \frac{k_{m}^{sr}}{P_{s}}\right)$,
and $\beta = \exp\left ( \frac{k_{m}^{sd}}{P_{s}}\right )E_{1}\left ( \frac{k_{m}^{sd}}{P_{s}}\right)$.
Setting $\nu_{m} = 0$ in \eqref{diffRes} and adding with \eqref{muzero} leads to $\nu_m+\mu_m = 0$. But neither $\nu_m$ or $\mu_m$ can be lesser than 0 due to the conditions $\nu_m\ge0$ and $\mu_m\ge0$. Thus the multipliers $\nu_m = \mu_m = 0$.
Equation~\eqref{muzero} with $\nu_m = 0$ is a quartic equation. To solve for $P_{m}$, we need to find the roots by first converting the regular quartic into a depressed quartic function of the form,
\begin{equation}\label{depquart}
P_{m}^4 + \lambda_1P_{m}^3 + \lambda_2P_{m}^2 + \lambda_3P_{m} - \lambda_4 = 0,
\end{equation}
\begin{equation}
\textrm{where,}\;\;\;\;\;\lambda_1 = \frac{-k_{m}^{rd}(2c_{m}k_{m}^{sd}-1)}{c_{m}} - \frac{1}{\tau \log e},
\end{equation}
\begin{equation}
\lambda_2 = \frac{k_m^{rd}}{c_{m}}\left[ 2k_m^{sd}(c_{m}k_m^{rd}k_m^{sd} -1)+\frac{k_m^{sd}[c_{m}(1-\beta)+b_{m}]}{\tau\log e} +\frac{k_m^{rd}[b_{m}-1]}{\tau\log e}\right],
\end{equation}
\begin{equation}
\lambda_3 = \frac{k_m^{rd}k_m^{sd}\left[(k_m^{rd})^2k_m^{sd}\tau\log e+b_{m}+(1-\beta)\right]}{c_{m}\tau\log e}
+\frac{(k_m^{rd})^2a_{m}}{c_{m}\tau\log e},
\end{equation}
\begin{equation}\label{lambda4}
\textrm{and}\;\;\;\;\lambda_4  = \frac{(k_m^{rd})^3k_m^{sd}a_{m}}{c_{m}\tau\log e}.
\end{equation}
$P_{m}$ is then, one of the roots to the quartic function \cite{Abrahamovitz} in \eqref{depquart}. The four roots of the quartic function are given by,
\begin{equation}\label{quartRoot}
\phi_m(d_m^{rd}, d_m^{sd}, P_{s}) =
\begin{cases}
\frac{\eta_{1} \pm \eta_2}{2}- \frac{\lambda_1}{4}\\
\frac{-\eta_1 \pm \eta_3}{2}- \frac{\lambda_1}{4}\\
\end{cases}
\end{equation}
\begin{equation}
\textrm{where, }\;\;\eta_{1}  = \sqrt{\frac{\lambda_1^2}{4}-\lambda_2+ \theta},\;\;\;\;\;
\eta_{2} =
\begin{cases}
\sqrt{\frac{3\lambda_1^2}{4}-\eta_1^2-2\lambda_2+ \frac{\left(4\lambda_1\lambda_2-8\lambda_3-\lambda_1^3\right)}{4\eta_1}} \;\;\;\;\;\; \text{if}\;\; \eta_1\neq0 \\
\sqrt{\frac{3\lambda_1^2}{4}-2\lambda_2+ \sqrt{\theta^2 - 4\lambda_4}} \;\;\;\;\;\;\;\;\;\;\;\;\;\;\;\;\;\;\; \text{if} \;\;\eta_1 = 0 ,\\
\end{cases}
\end{equation}
\begin{equation}
\textrm{and}\;\;\eta_{3} =
\begin{cases}
\sqrt{\frac{3\lambda_1^2}{4}-\eta_1^2-2\lambda_2 - \frac{\left(4\lambda_1\lambda_2-8\lambda_3-\lambda_1^3\right)}{4\eta_1}} \;\;\;\;\;\; \;\;\;\text{if}\;\; \eta_1\neq0 \\
\sqrt{\frac{3\lambda_1^2}{4}-2\lambda_2 - \sqrt{\theta^2 - 4\lambda_4}} \;\;\;\;\;\;\;\;\;\;\;\;\;\;\;\;\;\;\;\;\;\; \text{if} \;\;\eta_1 = 0 .\\
\end{cases}
\end{equation}
In the above equations, $\theta$ is defined as
\begin{equation}
\theta = \frac{\lambda_2}{3}-\frac{ \sqrt[3]{2} \left(-\lambda_2^2+3\lambda_1\lambda_3-12 \lambda_4\right)}{\sqrt[3]{\upsilon_1+\sqrt{\upsilon_2+\upsilon_1^2}}}+\frac{\sqrt[3]{\upsilon_1+\sqrt{\upsilon_2+\upsilon_1^2}}}{\sqrt[3]{32}},
\end{equation}
where, $\upsilon_1 = 2\lambda_2^3-9\lambda_1\lambda_2\lambda_3+27\lambda_3^2+27\lambda_1^2\lambda_4-72\lambda_2\lambda_4$, and
$\upsilon_2 = 4(3\lambda_1\lambda_3-\lambda_2^2-12\lambda_4)$.
Note that there are four possible roots for $\phi_m(d_m^{rd}, d_m^{sd}, P_{s})$ in \eqref{quartRoot}. A closer inspection reveals that not all the roots are useful. This is because in \eqref{depquart}, the coefficient $\lambda_4 $ defined in \eqref{lambda4} is always greater than 0, since $k_m^{rd}$, $k_m^{sd}$ and $\tau$ are greater than 0. From Descrates's sign rule \cite{Abrahamovitz}, there are either three roots or only one positive root for the quartic function in \eqref{depquart}, irrespective of the sign of $\lambda_1, \lambda_2$ or $\lambda_3$.
The first root in \eqref{quartRoot}, is positive, since, $\eta_1, \eta_2$ and $\eta_3$ are $\ge0$ and $\lambda_1 \leq 0$. When three positive roots are available, the second, third and fourth root in \eqref{quartRoot} are closer to or less than zero when the multiplier $\tau <1$. Since a small value of $\tau$ is desirable \cite{WeiYu}, only the first root in \eqref{quartRoot} is useful. $P_m$ is then given by,
\begin{equation}\label{quartRoot1}
P_m = \phi_m(d_m^{rd}, d_m^{sd}, P_{s}) =  \frac{\eta_{1}+ \eta_2}{2}- \frac{\lambda_1}{4}.
\end{equation}
The power allocation vector $P_{m}$ can be then be summarized as in \eqref{apprxpowth}. The optimal Lagrange's multiplier $\tau^*$, which maximizes \eqref{maxeq}, can be found through a one-dimensional search (e.g. using bisection \cite{bisection}), such that the constraint $\sum_{m = 1}^{M}\phi_m(d_m^{rd}, d_m^{sd}, P_{s}) = P_{r}$ is satisfied. Note that when all the sources are at equal distances to the relay and destination, $k_m^{sd}$, $k_m^{sr}$ and $k_m^{rd}$ are equal for all sources $ m = 1,...,M$. This then leads to the parameters $\lambda_1$, $\lambda_2$, $\lambda_3$, $\lambda_4$, $\eta_1$, $\eta_2$, $\eta_3$, $\theta$, $\upsilon_1$ and $\upsilon_2$ to be equal for all sources, for any value of the multiplier $\tau$. Due to the constraint, $\sum_{m = 1}^{M}\phi_m(d_m^{rd}, d_m^{sd}, P_{s}) = P_{r}$, equal power is allocated to all the sources.
\ifCLASSOPTIONcaptionsoff
\newpage \fi


\begin{table*}[h]
\caption{Rational function constants for different ranges of $\Delta_k, k \in \{1,2,3\}$}
\begin{center}
\begin{tabular*}{0.5\textwidth}{@{\extracolsep{\fill}} || c || c || c || c ||}
\hline
  $10\log_{10}(\Delta_k)$ in dB & $a$ & $b$ & $c$ \\
      \hline  $\Delta_{1} = \{-15~\textrm{to}~0\}$~dB  &2.4989  &0.0364  &0.005416 \\
     \hline  $\Delta_2= \{0~\textrm{to}~15\}$~dB   &0.3495  &0.3698  &0.0985 \\
     \hline  $\Delta_3= \{15~\textrm{to}~30\}$~dB &0.003246  &0.9306  &0.583 \\
    \hline
\end{tabular*}
\end{center}
\label{StatTable1}
\end{table*}

\begin{table*}[h]
\caption{Number of Computations Required for the PAS-1 and PAS-2 algorithms per iteration.}
\begin{center}
\begin{tabular*}{0.6\textwidth}{@{\extracolsep{\fill}} | c | c | c | }
  \hline $\textrm{Operation}$  &PAS-1 Algorithm &PAS-2 Algorithm\\
  \hline $\textrm{Mult.}$       &$M\log M + 64M$ &$M\log M + 10M$\\
  \hline $\textrm{Div.}$        &$M\log M + 15M$ &$M\log M + 3M$\\
  \hline $\log(\cdot)$          &$M\log M$       &0 \\
  \hline $\exp(\cdot)$          &$2M$            &$2M$  \\
  \hline $E_1(\cdot)$           &$2M$            &$2M$\\
  \hline $\sqrt[2](\cdot)$      &$8M$            &$M$ \\
  \hline $\sqrt[3](\cdot)$      &$2M$            &0\\
  \hline
\end{tabular*}
\end{center}
\label{StatTable2}
\end{table*}

\end{document}